# Measurements in Interstellar Space of Galactic Cosmic Ray Isotopes of

# Li, Be, B and N, Ne Nuclei Between 40-160 MeV/nuc by the

# CRS Instrument on Voyager 1

W.R. Webber[1], N. Lal[2] and B. Heikkila[2]

1. New Mexico State University, Astronomy Department, Las Cruces, NM  88003, USA

2. NASA/Goddard Space Flight Center, Greenbelt, MD 20771, USA

# ABSTRACT

In this paper we report a study of the isotopic composition of Li, Be, B and N, Ne nuclei from a 5 year time period beyond the heliopause using the CRS instruments on Voyager. By comparing the isotopic ratios, $^{15}N/^{14}N$ and $^{22}Ne/^{20}Ne$ outside the heliosphere as measured at Voyager, and which are found to be significantly lower than those measured at the same energy inside the heliosphere, we have provided strong evidence that cosmic rays of this energy have lost as much as 200 MeV/nuc or more in the solar modulation process. This is in accordance with the so called "force field" description of this overall modulation by Gleeson and Axford. The measurements at Voyager confirm that the unusual $^{14}N$ and $^{22}Ne$ cosmic ray source abundances relative to solar abundances made earlier inside the heliosphere extend to the lower energies not accessible from near Earth measurements. The low energy Li, Be and B nuclei, which are believed to be purely secondary nuclei, are found to have a (previously unobservable) peak in the differential intensity spectrum at ~100 MeV/nuc. This is in agreement with propagation predictions. The intensities of these nuclei are ~10-20% higher than those predicted in a propagation model with a matter path length $\lambda = 9$ g/cm$^2$ at these low energies. The isotopic composition of Li, Be and B nuclei is also consistent with that expected from propagation through interstellar matter.

## 1. Introduction

Measurements of the isotopes of galactic cosmic rays near the Earth have been very valuable in the effort to understand the sources of these energetic particles. Two of the most significant of these measurements have involved the isotopes $^{14}$N and $^{22}$Ne. The abundance of $^{14}$N in cosmic rays is found to be considerably under-abundant relative to the solar $^{14}$N/$^{16}$O abundance ratio which is 0.12 (Lukasiak, et al., 1994). In the case of $^{22}$Ne the opposite is true, $^{22}$Ne is significantly over-abundant relative to the solar $^{22}$Ne abundance. The cosmic ray $^{22}$Ne/$^{20}$Ne abundance ratio is ~5 times larger than the solar ratio (Binns, et al., 2002). These measurements made inside the heliosphere are subject to the effects of solar modulation which, among other things, cause the galactic cosmic rays to lose ~200 MeV/nuc or more in the heliosphere. Thus the composition of the lowest energy cosmic ray particles is not directly visible at the Earth.

Voyager 1 has now been beyond the heliopause (HP) for over 5 years measuring these low energy cosmic rays. Previous studies using the Voyager spacecraft over a 15 year time period within the heliosphere have provided a baseline for the work in this paper (Lukasiak, et al., 1994). The mass resolution for the Li, Be, B nuclei and the C, N, O nuclei is sufficient, for example, to resolve $^{13}$C from $^{12}$C even though the abundance of $^{13}$C is only ~1% of $^{12}$C.

## 2. The Voyager Measurement of Li, Be, B, N and Ne Mass Abundances

The HET telescope used in this analysis is shown and described in the above references. The isotopic abundances are obtained in a range from ~40 to 160 MeV/nuc. The mass analysis is made directly from the matrix of events, C432 vs. (B1+B2), where the pulse heights B1 and B2 are subject to the criteria: 1.30 < (B1/B2) < 2.16. The total energy C432 counter consists of 3-6 mm thick Lithium drift Si counters.

A sample of a pulse height matrix, C4 vs. (B1+B2), with this criteria for Li, Be and B nuclei is shown in Figure 1. The isotopes $^{6}$Li and $^{7}$Li are completely separated as are the isotopes $^{7}$Be and $^{9}$Be. There is one possible $^{10}$Be nucleus. The $^{10}$B and $^{11}$B isotopes are also completely resolved visually.

From these matrices mass lines are obtained for the various isotopes. The mass distribution for N nuclei from 61 to 149 MeV/nuc is shown in Figure 2. This is almost identical to the resolution obtained for N isotopes in the earlier studies of Lukasiak, et al., 1994, using a somewhat different technique for mass separation.

The results from this analysis for Li, Be, B, N and Ne nuclei are shown in Tables 1 and 2. For N nuclei the $^{15}N/^{14}N$ ratios obtained from this study and the earlier study at Voyager inside the heliosphere, but displaced upward in energy by ~200 MeV/nuc to account for solar modulation, are shown in Figure 3. For Ne, the $^{22}Ne/^{20}Ne$ and $^{21}Ne/^{20}Ne$ ratios from this study are shown in Figure 4, along with the ACE data from 1997-1998, also displaced upward by 200 MeV/nuc (Binns, et al., 2002). In both cases, if the inside the heliopause results were plotted at the measured energies, they would be 25-35% higher than those measured at Voyager.

### 3. Cosmic Ray Propagation in the Galaxy and Solar Modulation in the Heliosphere
#### a. N and Ne Isotope Ratios

The comparison of the N and Ne isotope ratios measured both inside and outside the heliosphere, which are shown in Figures 3 and 4, provide a valuable test of both solar modulation theories and galactic propagation features.

It is clear that the LIS $^{15}N/^{14}N$ and $^{22}Ne/^{20}Ne$ ratios measured at V1 beyond the heliopause are significantly lower than those measured within the heliosphere at the same energy. The energy range of each measurement inside or outside the heliosphere is identical, since, for both N and Ne nuclei they are made with the same instrument on Voyager.

The solid black lines in Figures 3 and 4 are from Leaky Box propagation model calculations in which the propagation parameters are "tuned" to fit the observed intensities and spectra of H and He, B, C and O nuclei from ~1 MeV/nuc to 1 TeV/nuc using Voyager data at lower energies and AMS-2 data above ~10 GV/nuc (Webber, et al., 2017; see also Webber, et al., 2018). These papers also provide a list of predicted intensities after interstellar propagation for Z=1 through 8 nuclei and for energies between 1 MeV and 1 TeV.

These propagation parameters include a path length $\lambda = 22.3 \, \beta \, P^{-0.45}$ in g/cm$^3$ above 0.88 GV (100 MeV/nuc for A/Z=2 particles; ~400 MeV for protons). Below 0.88 GV the path length is taken to be constant = 9 g/cm$^2$ down to a value of P ~0.3 GV.

The calculated LIS ratios, after propagation from the cosmic ray sources to heliosphere, for $^{15}$N/$^{14}$N and $^{22}$Ne/$^{20}$Ne respectively, are seen in Figures 3 and 4 to increase with increasing energy up to ~1 GeV/nuc. For the N isotopes this calculated increase matches the measured increase within and outside the heliosphere for a $^{14}$N/$^{16}$O abundance ratio of between 5-6% at the source. This matching with the LIS calculated ratio is only possible when the ratio measured within the heliosphere is moved to a ~200 MeV/nuc higher energy as would be expected for a modulation potential = 400 MV in the heliosphere (Gleeson and Axford, 1968; Webber, Stone, Cummings, et al., 2017).

The same is true for the $^{22}$Ne/$^{20}$Ne ratio, starting with a $^{22}$Ne/$^{20}$Ne source ratio = 0.395. The measured intensities inside the heliosphere and in LIS space both match the predictions, but only when the inside of the heliosphere energies are adjusted for an average energy loss ~200 MeV/nuc, corresponding to the modulation potential ~400 MV existing at the time of the ACE measurement. There is also some evidence from the lowest energy V1 data outside the heliosphere that the source ratio may decrease at low energies.

From both of these observations, however, there is a strong argument that the potential change that describes the solar modulation in the Gleeson and Axford model represents an actual average energy loss between the two locations inside and outside the heliosphere. Therefore particles with energies below ~200 MeV/nuc in interstellar space are not directly observable in the inner heliosphere.

**b.  Li, Be and B Nuclei**

For Li, Be and B, the observed LIS abundances are expected to be all secondaries, unlike the N and Ne nuclei. The calculated intensities of these nuclei after propagation have similar differential spectra and these spectra have a maximum at ~100 MeV/nuc.

Figure 5 shows the measured and the calculated intensities of Li, Be and B nuclei in the LIM obtained using the same propagation parameters as above. For these nuclei the measured and calculated intensities are generally within the experimental errors on the measurements and cross sections/path length used for the calculation. However, the measured absolute intensity measurements at Voyager appear to be 10-20% higher than the predictions in each case as can be seen in Figure 5. When more Voyager data is accumulated this data will provide a better test of the uncertainties of the cross sections and of the assumed path length used in the propagation model.

## **4. Summary and Conclusions**

As a result of our measurements of the $^{15}$N/$^{14}$N and $^{22}$Ne/$^{20}$Ne ratios at Voyager outside the heliopause we conclude that the effects of energy loss in solar modulation process in the heliosphere prevent galactic cosmic rays of less than ~200 MeV/nuc from reaching the inner heliosphere.

Cosmic ray Li, Be and B nuclei are purely secondary nuclei and have a maximum in their differential spectra at ~100 MeV/nuc after propagation, which is below this threshold. The intensity of these nuclei now observed directly for the 1$^{st}$ time at Voyager depends on their production cross sections and the amount of interstellar matter traversed. For the cross sections used in our program the observed abundances of each of these nuclei at Voyager are 10-20% higher than predictions using a matter path length ~9 g/cm$^2$. The measured isotopic abundance fractions of Li, Be and B nuclei are consistent with those obtained from the propagation program to within the statistical accuracy of the current Voyager data.

The isotopes $^{14}$N and $^{22}$Ne are two of the most interesting isotopes among the primary cosmic ray nuclei. Both of these isotopes have greatly different cosmic ray abundance ratios, e.g., $^{15}$N/$^{14}$N and $^{22}$Ne/$^{20}$Ne, than the solar ratios, for example.

The calculated LIS ratios of $^{15}$N/$^{14}$N and $^{22}$Ne/$^{20}$Ne are found to increase by 33% and 30% respectively between ~100 MeV/nuc, the average energy/nuc actually measured at Voyager, and ~300 MeV/nuc, the measured energy + 200 MeV/nuc to account for energy loss in the heliosphere. Since the average ratios for these isotopes measured at Voyager and ACE within the heliosphere, where the solar modulation potential is 400 MV, are indeed about 30% higher than those measured by Voyager 1, if the heliosphere measurements are displaced by +200

MeV/nuc then both sets of measurements inside and outside the heliosphere agree with the LIS calculated spectra.

This displacement of energy or average energy loss in the heliosphere which is necessary to produce the above agreement is predicted by modulation models that use Liouville's theorem which relates to the constancy of the particles energy density and momentum in phase space, as a basis for the modulation, such as the "force field" model of Gleeson and Axford, 1968.

This is the first direct observation of this process working on a global scale in the outer heliosphere.

The Voyager measurements of these ratios are consistent with those measured earlier within the heliosphere if these considerations regarding the solar modulation are made. There is also some evidence from the Voyager data that the source $^{22}Ne/^{20}Ne$ ratio may decrease at LIS energies below ~200-300 MeV/nuc.

**Acknowledgments:** The authors are grateful to the Voyager team that designed and built the CRS experiment with the hope that one day it would measure the galactic spectra of nuclei and electrons. This includes the present team with Ed Stone as PI, Alan Cummings, Nand Lal and Bryant Heikkila, and to others who are no longer members of the team, F.B. McDonald and R.E. Vogt. Their prescience will not be forgotten. This work has been supported throughout the more than 40 years since the launch of the Voyagers by the JPL.

| TABLE I |
|---|
| B Stopping – 5 yr AVG |

| C4 Channel # | E-Loss (MeV) | Energy (MeV/nuc) | Cnts | Cnts/MeV | Int | GF x t | (x GF x t)$^{-1}$ |
|---|---|---|---|---|---|---|---|
| **$^{11}$B** | | | | | | | |
| 40 | 193 | 48.0 | | | | | |
| 80 | 186 | 58.0 | 105 | 10.5 | 1.030 | 2.72 | 3.97 |
| 130 | 626 | 74.4 | 152 | 9.26 | 1.050 | 2.44 | 3.98 |
| 184 | 893 | 95.1 | 163 | 7.87 | 1.062 | 2.15 | 3.89 |
| 240 | 1160 | 117.2 | 130 | 5.88 | 1.088 | 1.75 | 3.66 |
| **$^{10}$B** | | | | | | | |
| 37.5 | 185 | 50.6 | | | | | |
| 75 | 370 | 61.2 | 44 | 4.15 | 1.030 | 2.72 | 1.57 |
| 122 | 599 | 78.5 | 75 | 4.33 | 1.050 | 2.44 | 1.87 |
| 172 | 854 | 100.3 | 83 | 3.81 | 1.062 | 2.15 | 1.88 |
| 226 | 1110 | 123.0 | 54 | 2.38 | 1.078 | 1.75 | 1.47 |
| **$^{7}$Be** | | | | | | | |
| 25.4 | 121.5 | 48 | | | | | |
| 50.7 | 243 | 58 | 26 | 2.60 | 1.027 | 2.72 | 0.982 |
| 82 | 398 | 74.4 | 49 | 3.00 | 1.040 | 2.44 | 1.27 |
| 117 | 597 | 95.2 | 57 | 2.74 | 1.062 | 2.15 | 1.35 |
| 152 | 737 | 117.2 | 48 | 2.18 | 1.075 | 1.75 | 1.34 |
| **$^{9}$Be** | | | | | | | |
| 28.3 | 137 | 41.7 | | | | | |
| 56.5 | 274 | 50.4 | 7.5 | 0.86 | 1.030 | 2.72 | 0.325 |
| 92 | 444 | 64.6 | 9 | 0.64 | 1.048 | 2.44 | 0.28 |
| 131 | 633 | 82.6 | 17 | 0.944 | 1.060 | 2.15 | 0.47 |
| 170 | 822 | 101.5 | 14 | 0.74 | 1.082 | 1.75 | 0.45 |
| **$^{6}$Li** | | | | | | | |
| 17.3 | 81.7 | 37.9 | | | | | |
| 34.7 | 163.3 | 45.3 | 35 | 4.73 | 1.028 | 2.72 | 1.79 |
| 56.2 | 264.5 | 58.1 | 57 | 4.45 | 1.037 | 2.44 | 1.89 |
| 80 | 377 | 74.0 | 46 | 2.89 | 1.053 | 2.15 | 1.42 |
| 104 | 490 | 91.6 | 39 | 2.22 | 1.069 | 1.75 | 1.36 |
| **$^{7}$Li** | | | | | | | |
| 18.3 | 88.3 | 34.8 | | | | | |
| 36.6 | 176.6 | 42.0 | 16 | 2.22 | 1.028 | 2.72 | 0.84 |
| 59.4 | 28.8 | 53.6 | 32 | 2.76 | 1.037 | 2.44 | 1.17 |
| 84.7 | 40.8 | 68.5 | 27 | 1.82 | 1.053 | 2.15 | 0.89 |
| 110 | 53.0 | 84.4 | 24 | 1.91 | 1.069 | 1.75 | 0.97 |

| TABLE II | | | | | | | |
|---|---|---|---|---|---|---|---|
| B Stopping – 5 yr AVG | | | | | | | |
| <sup>20</sup>Ne | | | | | | | |
| C4 Channel # | E-Loss (MeV) | Energy (MeV/nuc) | Cnts | Cnts/MeV | Int | GF x t | (x GF x t)<sup>-1</sup> |
| 116 | 548.0 | 74.6 | | | | | |
| 234 | 1095.0 | 90.5 | 111 | 6.98 | 1.038 | 2.72 | 2.66 |
| 376 | 1775.0 | 116.2 | 135 | 5.25 | 1.065 | 2.44 | 2.29 |
| 536 | 2530.0 | 149 | 156 | 4.76 | 1.082 | 2.15 | 2.39 |
| 696 | 3290.0 | 183 | 111 | 3.26 | 1.118 | 1.75 | 2.09 |
| <sup>21</sup>Ne | | | | | | | |
| 117.5 | 567.0 | 72.8 | | | | | |
| 236 | 1132.0 | 88.6 | 8 | 0.51 | 1.038 | 2.72 | 0.193 |
| 381 | 1838.0 | 114.1 | 17 | 0.67 | 1.065 | 2.44 | 0.291 |
| 542 | 2580.0 | 145 | 20 | 0.65 | 1.082 | 2.15 | 0.326 |
| 704 | 3365.0 | 178 | 16 | 0.485 | 1.118 | 1.75 | 0.310 |
| <sup>22</sup>Ne | | | | | | | |
| 119 | 586.0 | 71.0 | | | | | |
| 238 | 1170.0 | 86.6 | 34 | 2.18 | 1.038 | 2.72 | 0.832 |
| 386 | 1900.0 | 112.1 | 61 | 2.39 | 1.065 | 2.44 | 1.04 |
| 548 | 2630.0 | 140.7 | 67 | 2.34 | 1.082 | 2.15 | 1.17 |
| 712 | 3420.0 | 173 | 52 | 1.61 | 1.118 | 1.75 | 1.03 |
| <sup>14</sup>N | | | | | | | |
| 66 | 3130.0 | 61 | | | | | |
| 132 | 624.0 | 74 | 119 | 9.14 | 1.033 | 2.72 | 3.53 |
| 214 | 1015.0 | 95.0 | 180 | 8.55 | 1.050 | 2.44 | 3.68 |
| 305 | 1450.0 | 121.3 | 194 | 7.38 | 1.070 | 2.15 | 3.68 |
| 396 | 1850.0 | 149 | 138 | 4.96 | 1.110 | 1.75 | 3.14 |
| <sup>15</sup>N | | | | | | | |
| 67 | 320.0 | 58.6 | | | | | |
| 134 | 640.0 | 70.9 | 80 | 6.50 | 1.033 | 2.72 | 2.47 |
| 217 | 1037.0 | 90.8 | 122 | 6.13 | 1.050 | 2.44 | 2.63 |
| 309.5 | 1480.0 | 116.0 | 133 | 5.28 | 1.070 | 2.15 | 2.63 |
| 402 | 1920.0 | 143.0 | 125 | 4.63 | 1.110 | 1.75 | 2.94 |

**FIGURE CAPTIONS**

**Figure 1:**  A matrix of the C432 vs. (B1+B2) pulse heights for Li, Be and B.  The isotopes of $^6$Li and $^7$Li (lower left), $^7$Be and $^9$Be and $^{10}$B and $^{11}$B (upper right) are seen to be well separated.

**Figure 2:**  The mass distribution of N isotopes from ~61 to 149 MeV/nuc.

**Figure 3:**  The calculated LIS $^{15}$N/$^{14}$N ratio as a function of energy using the LBM propagation described in the text.  The measurements of this ratio at Voyager beyond the heliosphere are shown.  The measurements of this ratio at Voyager inside the heliosphere are also shown (Lukasiak, et al., 1994).  The measurements of the ratio within the heliosphere are moved upwards in energy by 200 MeV/nuc to account for solar modulation.

**Figure 4:**  The calculated LIS $^{22}$Ne/$^{20}$Ne ratio as a function of energy.  The measurements of this ratio at Voyager 1 beyond the heliosphere are shown.  The measurements of the ratio within the heliosphere by Binns, et al., 2002, are also shown and are moved to a 200 MeV higher energy to account for the effects of the 400 MV solar modulation potential at the time of the measurements.

**Figure 5:**  The calculated and measured LIS intensities of Li, Be and B nuclei below ~200 MeV/nuc.  The path length in a LBM propagation is taken to be ~22.3 β P$^{-0.45}$ above 1 GV and 9 g/cm below 1 GV as in the calculations in Figures 3 and 4.

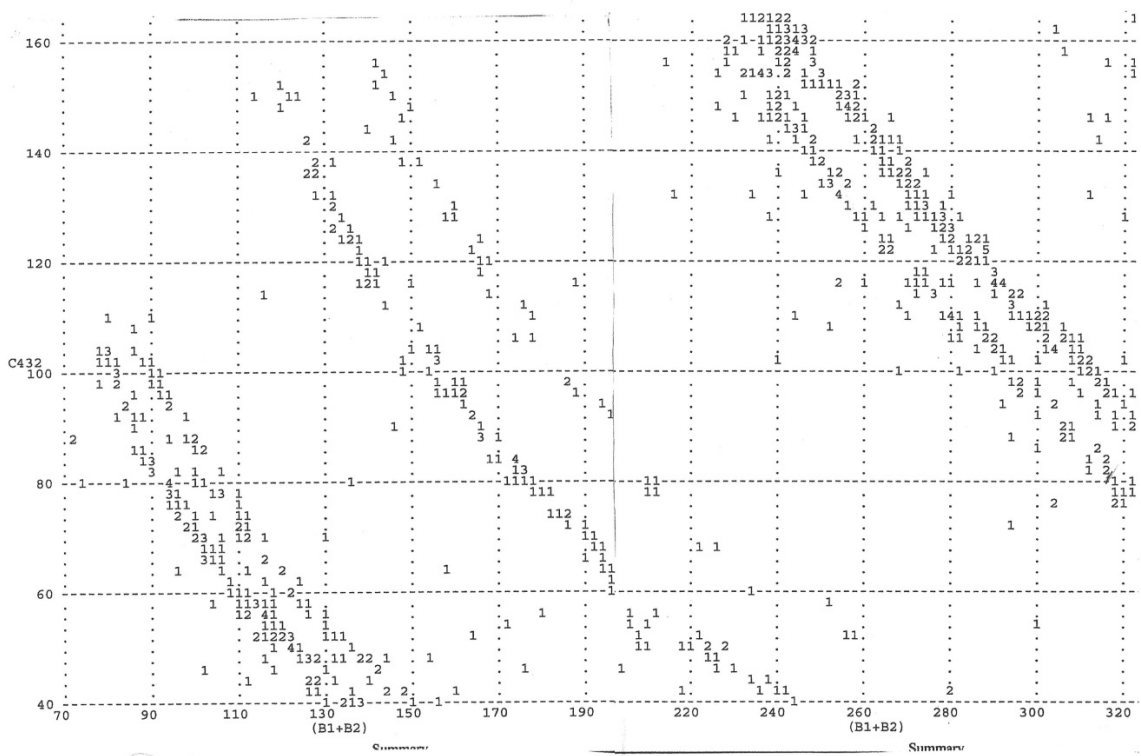

**FIGURE 1**

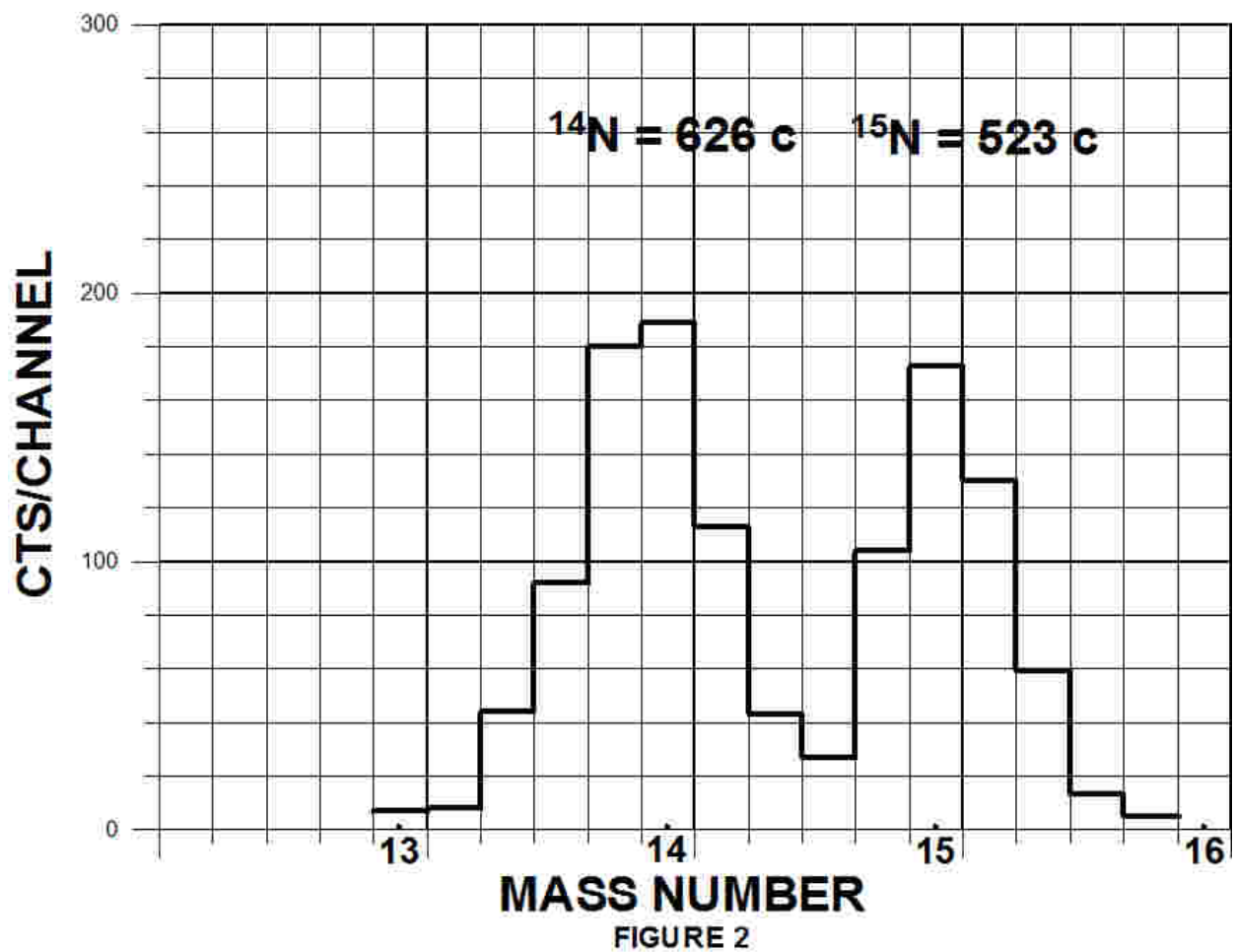

**FIGURE 2**

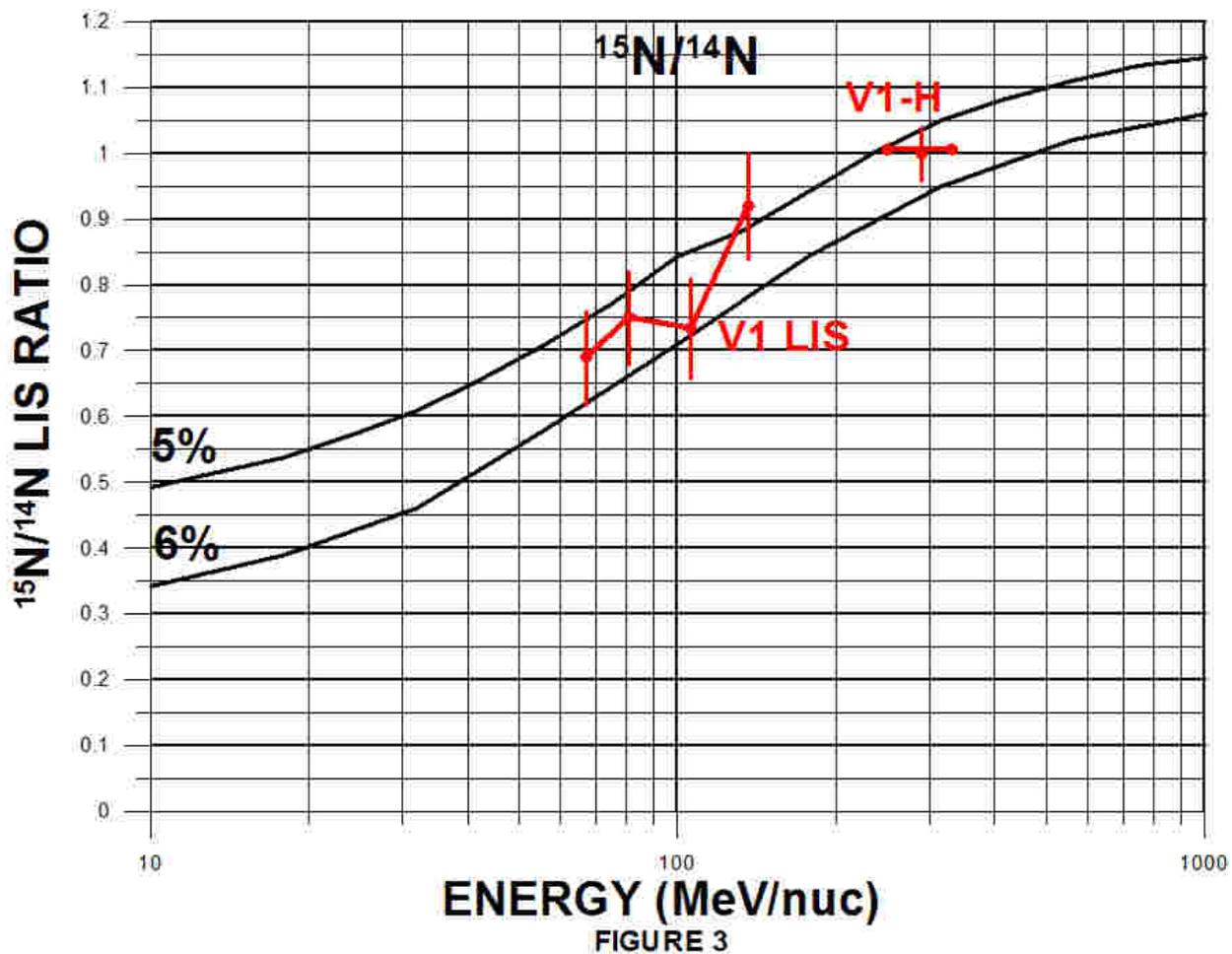

FIGURE 3

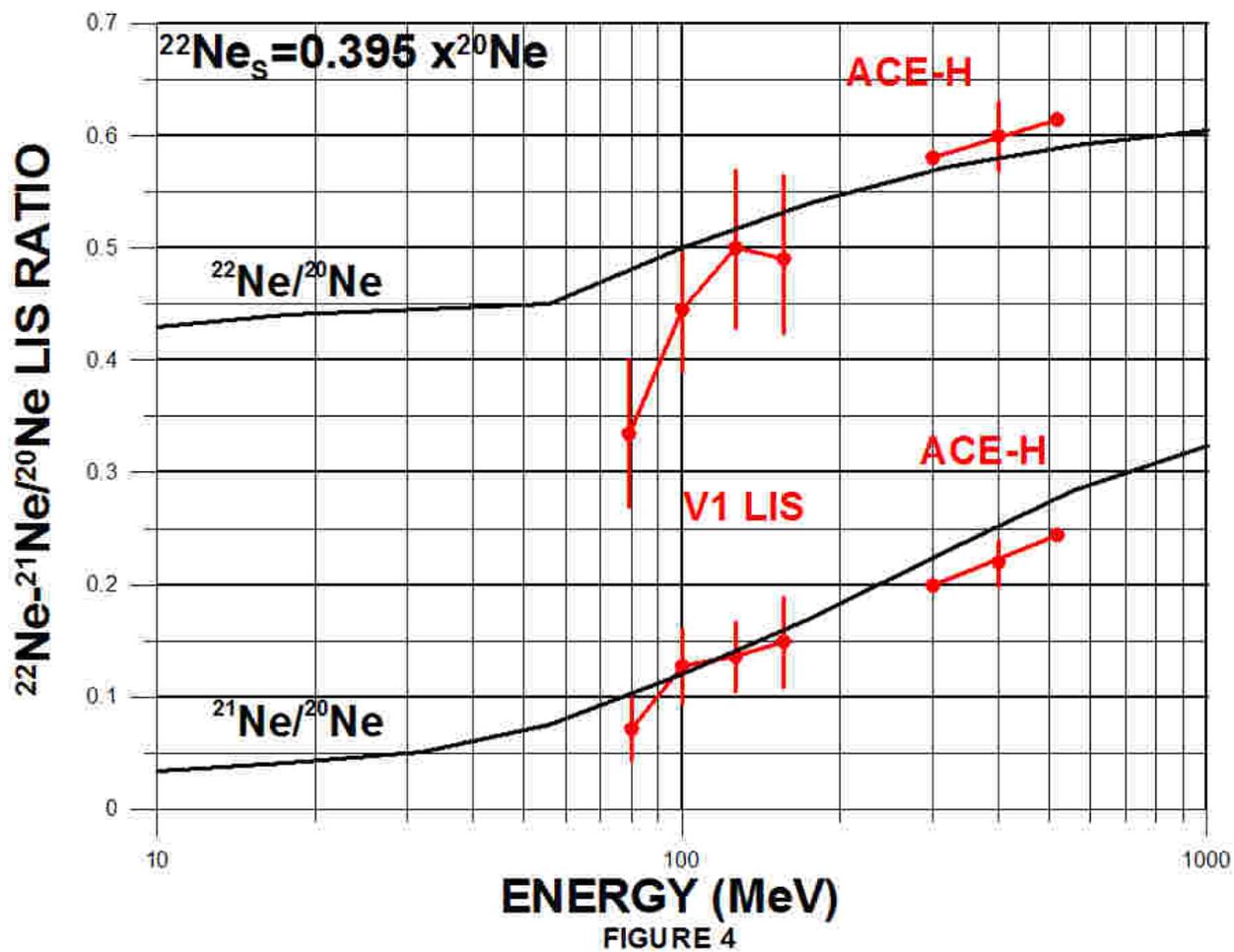

FIGURE 4

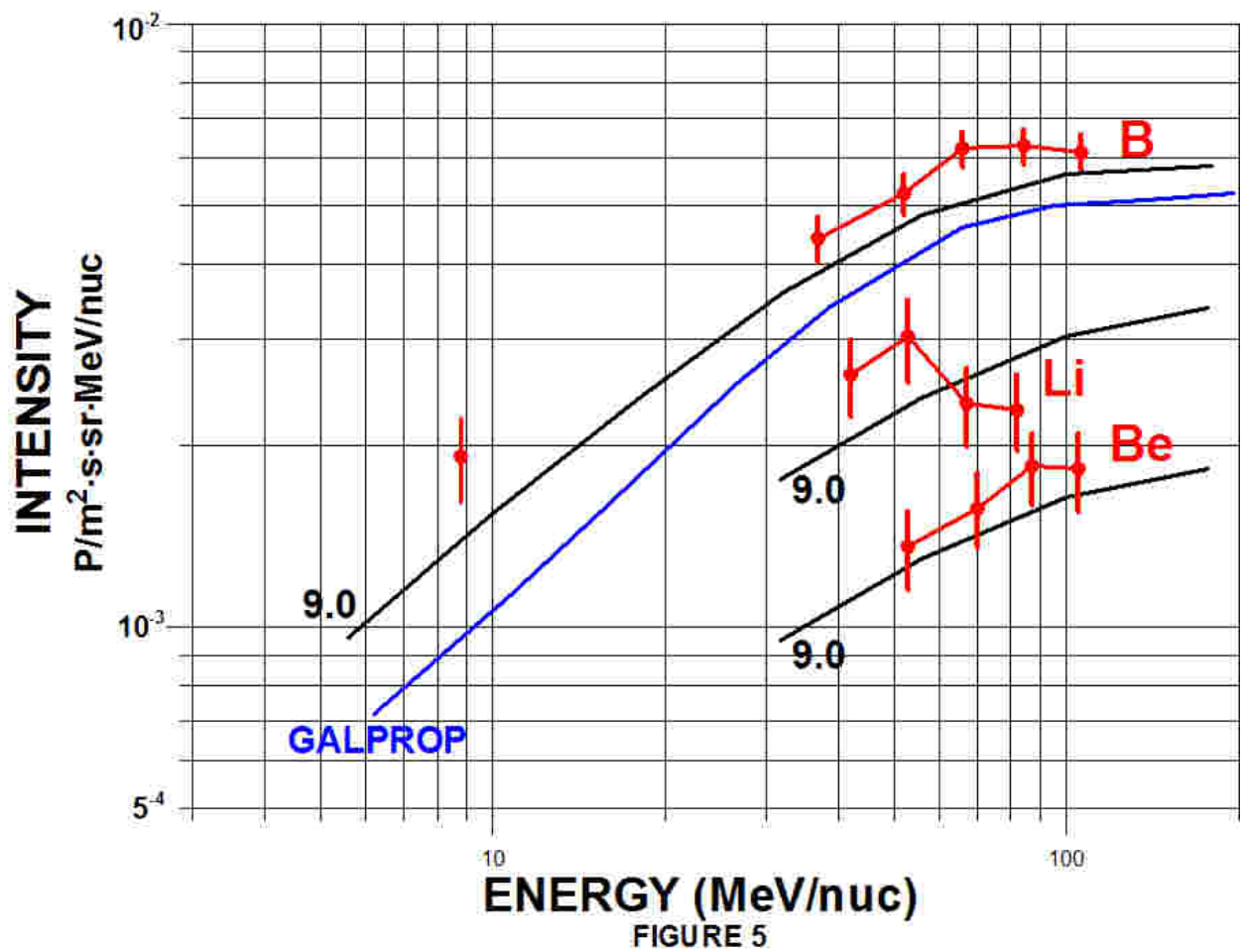

**FIGURE 5**